\documentclass[aps,pre,twocolumn,groupedaddress,showkeys,showpacs,amsmath,amssymb,preprintnumbers]{revtex4}

\usepackage{graphicx}

\begin{document}

\title{Stabilisation of the lattice-Boltzmann method \\ using the Ehrenfests' coarse-graining}
\author{R. A. Brownlee}
\email[corresponding author: ]{r.brownlee@mcs.le.ac.uk}
\author{A. N. Gorban}
\author{J. Levesley}
\affiliation{Department of Mathematics, University of Leicester,
Leicester LE1 7RH, UK}

\date{\today}

\begin{abstract}
The lattice-Boltzmann method (LBM) and its variants have emerged as
promising, computationally efficient and increasingly popular
numerical methods for modelling complex fluid flow. However, it is
acknowledged that the method can demonstrate numerical
instabilities, e.g., in the vicinity of shocks. We propose a simple
and novel technique to stabilise the lattice-Boltzmann method by
monitoring the difference between microscopic and macroscopic
entropy. Populations are returned to their equilibrium states if a
threshold value is exceeded. We coin the name \textit{Ehrenfests'
steps} for this procedure in homage to the vehicle that we use to
introduce the procedure, namely, the Ehrenfests' idea of
coarse-graining.
\end{abstract}



\maketitle

\section{Introduction}\label{sec1}

The lattice-Boltzmann method (LBM) provides an alternative to the
orthodox approach to computational fluid dynamics, in which the
starting point is always a discretization of the Navier--Stokes
equations. The method, which is fundamentally based on Boltzmann's
kinetic transport equation, instead describes a fluid by a number of
interacting populations of particles moving and colliding on a fixed
lattice. With the advent of the introduction of a diagonal collision
integral with single-time relaxation to local equilibrium, the
method becomes tantalisingly simple and efficient.

In recent years, the LBM has enjoyed much applied success modelling
various flows of genuine engineering interest (see
e.g.,~\cite{succi01} and the references therein). However, when
populations are far from equilibrium, such as is the case in the
vicinity of shocks, the LBM exhibits numerical instability. Often,
numerical instability in the LBM is attributed to the absence of a
positivity constraint on the populations. Recent development of the
entropic LBM (ELBM)~\cite{karlin00,boghosian01,karlin02} are
attempts to improve stability properties through compliance with a
discrete entropy $H$-theorem. Although such techniques should,
conceivably, reduce the magnitude of spurious oscillations, one
should perhaps not be disappointed to find that such numerical
instabilities are not removed entirely.

In this paper we suggest an alternative and versatile approach. The
idea is simply stated: we propose a LBM in which the difference
between microscopic and macroscopic entropy is monitored in the
simulation, and populations are returned to their equilibrium states
if a threshold value is exceeded. This technique is appealing
because dissipation is introduced in a controlled, targeted and
illiberal manner. Furthermore, we stress that equilibration itself
will leave macroscopic entropy completely unchanged.

We coin the name \textit{Ehrenfests' steps} for this procedure
because we feel that the technique is most clearly understood if
introduced using the Ehrenfests' coarse-graining
idea~\cite{ehrenfest11}.

This report is organised as follows. In Section~\ref{sec2} the LBM
is recalled. In Section~\ref{sec3} we recall the basic theory
required to discuss the Ehrenfests' coarse-graining and introduce
Ehrenfests' steps. The reader is directed to~\cite{gorban06} for a
more comprehensive review of coarse-graining. Finally, we present
the results of a shock tube numerical experiment which compares
various LBMs.

\section{Lattice-Boltzmann method}\label{sec2}

The Boltzmann kinetic transport equation is the following time
evolution equation for one-particle distribution functions
$f=f(x,v,t)$:
\begin{equation*}
    \partial_t f+v \cdot \nabla f = Q.
\end{equation*}
The \textit{collision
integral}, $Q$, describes the interactions of the populations $f$.
The lattice-Boltzmann approach drastically simplifies this model by
stipulating that populations can only move with a finite number of
velocities $\{v_1,\ldots,v_n\}$:
\begin{equation}\label{lbe}
    \partial_t f_i+v_i \cdot \nabla f_i  =
    Q_i,\quad \text{$i=1,\ldots,n$,}
\end{equation}
where $f_i$ is the one-particle distribution function associated
with motion in the $i$th direction. For example, in one-dimension,
one might consider three velocities $\{-c,0,c\}$, for some $c\neq
0$.

The model is further simplified by specifying the collision integral
as the Bhatnager--Gross--Krook (BGK) operator~\cite{bgk54}: $Q_i =
(f_i^\mathrm{eq}-f_i)/\tau$. We make this assumption for the
remainder of the paper. Now,~\eqref{lbe} describes free-flight
dynamics together with relaxation to local equilibria,
$f_i^{\mathrm{eq}}$, in time proportional to $\tau$.

The discrete velocities and local equilibrium states can all be
mindfully chosen so that the Navier--Stokes equations are recovered
by the lattice-Boltzmann equation~\eqref{lbe}, subject to ceratin
conservation laws, in the large time-scale, $t$, limit via a
Chapman--Enskog procedure~\cite{succi01}. Here, the macroscopic
fluid density $\rho$ and momentum $\rho u$ are the zeroth- and
first-order hydrodynamic moments of the populations respectively.
The rate of dissipation introduced by the model is proportional to
$\tau$.

The local equilibrium states can be found by maximising a local
entropy functional subject to the constraints of conservation of
mass and momentum.

There are numerous ways to discretize~\eqref{lbe} and obtain a
numerical method. We prefer the following, second-order accurate in
time, lattice-based LBGK scheme~\cite{succi01}:
\begin{equation}\label{lbm}
  f_i(x+v_i \Delta t,t+\Delta t) = (1-\beta)f_i(x,t)+\beta \tilde{f}_i(x,t),
\end{equation}
with $\tilde{f}_i = 2 f^\mathrm{eq}_i-f_i$. Here, the discrete
velocities are associated with an underlying spatial lattice
$\mathcal{L}$ and $x \in \mathcal{L}$. Populations live on this
lattice, propagate to neighbouring lattice sites with their
corresponding discrete velocities and are updated via~\eqref{lbm}.
The parameter $\beta \in (0,1]$ controls the viscosity in the model,
with $\beta=1$ corresponding to the zero-viscosity limit.

\section{The Ehrenfests' coarse-graining\label{sec3}}

To introduce the Ehrenfests' coarse-graining idea~\cite{ehrenfest11}
we use a formal kinetic equation
\begin{equation}\label{kinetic_eqn}
    \frac{\mathrm{d}f}{\mathrm{d}t} = J(f),
\end{equation}
together with a strictly concave entropy functional $S$. We tacitly
assume that entropy does not decrease with time:
$\mathrm{d}S(f)/\mathrm{d}t \geq  0$. For our purposes, we are
always interested in the example of the lattice-Boltzmann
equation~\eqref{lbe} with associated entropy functional which
defines the local equilibrium states.

The Ehrenfests' idea was to supplement the mechanical motion
from~\eqref{kinetic_eqn} with periodic averaging in cells to produce
piecewise constant, or \textit{coarse-grained}, densities. This
operation necessitates entropy production.

We wish to allow a generalisation of the Ehrenfests' coarse-graining
whereby averaging in cells is replaced with some other partial
equilibration procedure. Specifically, we assume we have some linear
operator $m$ which transforms a microscopic description of the
system $f$ into a macroscopic description $M=m(f)$. For our example,
the macroscopic description is that provided by the usual
hydrodynamic moments. Now, given a macroscopic description, $M$, we
consider the solution $f^*_M$ of the optimisation problem
\begin{equation}\label{opti_problem}
    \arg\max \Bigl\{ S(f): m(f)=M \Bigr\}.
\end{equation}

Averaging in cells is a particular example of this entropy
maximisation problem for the Boltzmann--Gibbs--Shannon (BGS) entropy
functional $S(f) = -\int f \log(f)$, where the integration is taken
over the whole of phase space.

We will refer to $f^*_M$ as the quasi-equilibrium distribution. The
quasi-equilibrium manifold $Q$ is the set of quasi-equilibrium
distributions parameterised by the macroscopic variables $M$.

\subsection{The Ehrenfests' chain and entropic involution\label{sec3-1}}

Entropy maximisation leads naturally to an evolution equation for
the macroscopic description. The so-called \textit{quasi-equilibrium
approximation} to~\eqref{kinetic_eqn} is an equation for the
evolution of $M$:
\begin{equation}\label{qe-approx}
    \frac{\mathrm{d}M}{\mathrm{d}t} = m\bigl(J(f^*_M)\bigr).
\end{equation}
For our example, the quasi-equilibrium distributions are precisely
the local equilibria $f^\mathrm{eq}_i$ and~\eqref{qe-approx}
coincides with the compressible Euler equations~\cite{gorban01}.

Now, one can envisage constructing various \textit{coarse-graining
chains} which provide stepwise approximation to the macroscopic
equations~\eqref{qe-approx}.

Let $\Theta_t$ be the phase flow for the kinetic
equation~\eqref{kinetic_eqn}. Let $\tau$ be a fixed
\textit{coarse-graining time} and suppose we have an initial
quasi-equilibrium distribution $f_0$. The Ehrenfests' chain is the
sequence of quasi-equilibrium distributions $f_0, f_1, \ldots$,
where $f_i := f^*_{m(\Theta_\tau(f_{i-1}))}$.

Entropy increases in the Ehrenfests' chain. The entropy gain in a
link in the chain is made up from two parts: the entropy gain from
the mechanical motion (from $f_i$ to $\Theta_\tau(f_i)$) and the
gain from the equilibration (from $\Theta_\tau(f_i)$ to $f_{i+1}$).
Consequently, conservative systems become dissipative and
dissipative systems more so. The gain in macroscopic entropy in a
particular link is given by the expression
$S(f^*_{m(f_{i+1})})-S(f^*_{m(f_{i})})$. Note that there is zero
gain in macroscopic entropy from the equilibration part.

The Ehrenfests' chain provides a stepwise approximation to a
solution of some coarse-grained macroscopic equations via $M_i :=
m(f_i)$, $i=1,2,\ldots$. For our example, these equations are the
compressible Navier--Stokes equations~\cite{gorban01}. However, this
chain is computationally prohibitive because the rate of introduced
dissipation is proportional to $\tau$. The Ehrenfests' chain
corresponds to the following LBM:
\begin{equation*}\label{lbm-euler}
  f_i(x+v_i \Delta t,t+\Delta t) = (1-\beta)f_i(x,t)+\beta {f}_i^{\mathrm{eq}}(x,t),
\end{equation*}
which we recognise as a forward Euler discretization of~\eqref{lbe}.

Another possibility is to construct a chain as follows. As already
noted, the dissipative term introduced by the Ehrenfests' chain
depends linearly on $\tau$. Therefore, there is symmetry between
forward and backward motion in time starting from any
quasi-equilibrium initial condition. It is precisely this principle
that enables one to construct chains with zero macroscopic entropy
production. Each subsequent link in the chain is constructed using
entropic involution.

To make such a chain useful, the user is at liberty to add a
required amount of dissipation by shifting the involuted point in
the direction of the quasi-equilibrium state, with some entropy
increase. Of course, this shift leaves macroscopic entropy
unchanged. This chain corresponds to the ELBM that we have eluded to
in the introduction:
\begin{equation}\label{elbm}
  f_i(x+v_i \Delta t,t+\Delta t) = (1-\beta)f_i(x,t)+\beta \tilde{f}_{i,\alpha}(x,t),
\end{equation}
with $\tilde{f}_{i,\alpha} = (1-\alpha) f_i+\alpha f^\mathrm{eq}_i$.
The number $\alpha=\alpha(f_i)$ is chosen so that a constant entropy
condition is satisfied. This, in itself, provides a positivity
constraint on the populations.

If the entropic involution operator is partially linearised so that
subsequent links are constructed with the use of reflections in the
quasi-equilibrium manifold, rather than inversions, then the
corresponding LBM is precisely LBGK~\eqref{lbm}.

\subsection{Ehrenfests' steps\label{sec3-2}}

We can now state the main idea of the paper. We propose to create a
new chain. This chain begins and proceeds, for the bulk of time, as
the either the entropic involution chain or its linearised version,
as described in the previous subsection. However, the difference
between microscopic and macroscopic entropy is monitored throughout
the simulation, by which we mean the quantities
\begin{equation*}
\Delta S_i := S(f^*_{m(f_i)})-S(f_i).
\end{equation*}
A threshold value is set and an alarm is triggered if exceeded. The
alarm simply signals that a link from the Ehrenfests' chain
--- an Ehrenfests' step --- be used in place of a regular link of
the primary chain at this point. Links in the chain which are
intolerably far from their quasi-equilibrium states are merely
returned to their equilibrium. The result is a chain which provides
additional dissipation only where it is anticipated to be required.

Coupling the entropic involution chain with Ehrenfests' steps will,
in general, no longer constitute a chain with zero macroscopic
entropy production. Indeed, in a chain of length $N+1$ the total
macroscopic entropy gain is given by the expression $\sum_{i\in I}
\bigl[ S(f^*_{m(f_{i+1})})-S(f^*_{m(f_i)}) \bigr]$, where $I \subset
\{1,\ldots,N\}$ is the set of indices corresponding to the
Ehrenfests' steps.

\section{Numerical experiment\label{sec4}}

To conclude this report, we perform a one-dimensional numerical
experiment to demonstrate the performance of the proposed LBM
corresponding to a chain with Ehrenfests' steps. The implementation
is as follows. If $\Delta S_i > \delta$, for some given tolerance
$\delta>0$ then we accept an Ehrenfests' step. The resulting LBM is
as follows:
\begin{equation}\label{lbmes}
    f_i(x+v_i \Delta t,t+\Delta t) = \left\{
    \begin{aligned}
        &(1-\beta)f_i+\beta f^\mathrm{eq}_i,&& \text{$\Delta S_i > \delta$,} \\
        &(1-\beta)f_i+\beta \tilde{f}_{i},&\quad& \text{otherwise.}
    \end{aligned}\right.
\end{equation}
We have selected LBGK~\eqref{lbm} as the primary chain and we will
henceforth refer to this LBM as LBGK-ES.

The method LBGK-ES, as described, is a second-order accurate in
time scheme with first-order degradation in regions where
Ehrenfests' steps are employed. However, as is often found, when
the thickness of such regions is of the order of the lattice
spacing, the method remains second-order everywhere.

For contrast we are interested in comparing LBGK-ES with
LBGK~\eqref{lbe} and ELBM~\eqref{elbm} (as described
in~\cite{karlin02}). In all of our simulations we select the
three-velocity model mentioned in the introduction and a uniformly
spaced lattice. Further, we always choose $c=1$ and $\Delta t=1$.
The coefficient $\beta$, which controls the viscosity in the model,
is fixed at $\beta=1-10^{-9}$, which is close to the zero viscosity
limit. In each case the entropy is $S=-H$, with $H = f_1
\log(f_1/4)+f_2\log(f_2)+f_3 \log(f_3)$~\cite{karlin99}, where
$f_1$, $f_2$ and $f_3$ denote the static, left-moving and
right-moving populations respectively. For this entropy, the local
equilibria are available analytically for the three-velocity model.
They are given by the expressions
\begin{align*}
  f_1^{\mathrm{eq}} &= 2\rho \bigl(2-\sqrt{1+3 u^2}\bigr)/3,\\
  f_2^{\mathrm{eq}} &= \rho \bigl((3u-1)+2\sqrt{1+3 u^2} \bigr)/6,\\
  f_3^{\mathrm{eq}} &= \rho \bigl(-(3u+1)+2\sqrt{1+3 u^2} \bigr)/6.
\end{align*}
For LBGK-ES we fix the tolerance, $\delta$, to either $\delta =
10^{-3}$ or $\delta = 10^{-5}$.

As already mentioned, for ELBM, there is a parameter, $\alpha$,
which is chosen to satisfy a constant entropy condition. This
involves finding the nontrivial root of the equation
\begin{equation}\label{entropy_estimate}
  H(f+\alpha Q) = H(f).
\end{equation}
Inaccuracy in the solution of this equation can introduce artificial
viscosity. To solve~\eqref{entropy_estimate} numerically we employ a
robust routine based on bisection. The root is solved to an accuracy
of $10^{-15}$. Furthermore, we always ensure that the returned value
of $\alpha$ does not lead to an entropy decrease. So that our
results can be faithfully reproduced, we stipulate that we will
consider two possibilities if no nontrivial root
of~\eqref{entropy_estimate} exists. We either elect to select
$\alpha=1$ or we select the positivity bound $\alpha=\alpha_+$, with
$\alpha_+:= \min_{Q_i<0}|f_i/Q_i|$. For fairness, since both of
these procedures introduce diffusivity into ELBM, we consider
analogous procedures for LBGK too, i.e., we elect to select either
$\alpha=1$ or $\alpha = \alpha_+$ if a populations is predicted to
become negative.

\begin{figure}
\vspace{-0.75cm}
\includegraphics[height=14.0cm]{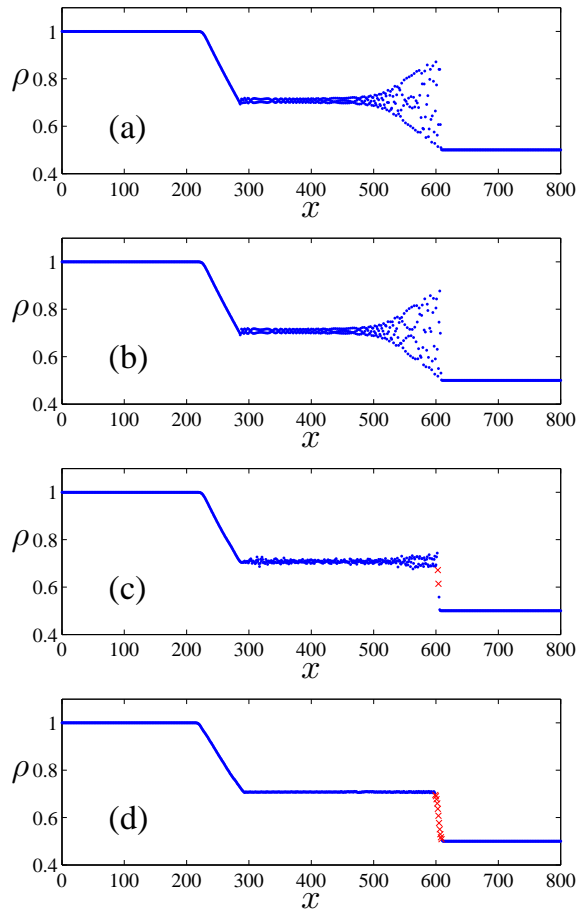}
\vspace{-1.0cm} \caption{Density profile of the isothermal $1:2$
shock tube simulation after $300$ time steps using
    (a) LBGK~\eqref{lbm};
    (b) ELBM~\eqref{elbm};
    (c) LBGK-ES~\eqref{lbmes} with $\delta=10^{-3}$;
    (d) LBGK-ES~\eqref{lbmes} with $\delta=10^{-5}$.
In this example, LBGK does not produce a negative population so the
aforementioned regularisation procedures are redundant. Similarly,
for ELBM in this example, the entropy estimate equation always has a
nontrivial root. Sites where Ehrenfests' steps are employed are
indicated by crosses. \label{shocktube}}
\end{figure}

\subsection{Shock tube results\label{sec4-1}}

The one-dimensional shock tube for a compressible isothermal fluid
is a standard benchmark test for hydrodynamic codes. Our
computational domain will be the interval $[0,800]$ and we
discretize this interval with $801$ uniformly spaced lattice
sites. We choose the initial density ratio as $1:2$. Initially,
for $x\leq 400$ we set $\rho=1.0$ else we set $\rho=0.5$.

We observe that, of all the LBMs considered in the experiment,
only the method which includes Ehrenfests' steps is capable of
suppressing spurious post-shock oscillations
(Fig.~\ref{shocktube}).

The code used to produce the simulations in this section is freely
available by making contact with the corresponding author.

\section*{Conclusion}
Ehrenfests' steps introduce additional dissipation \textit{locally},
on the base of \textit{point-wise analysis of non-equilibrium
entropy}. They admit a huge variety of generalisations: incomplete
Ehrenfests' steps, partial involution, etc.~\cite{gorban06}. In
order to preserve the second-order of LBM accuracy it is worthwhile
to perform Ehrenfests' steps on only a small share of sites (the
number of sites should be $\mathcal{O}(Nh/\ell)$, where $N$ is the
number of sites, $\ell$ is the macroscopic characteristic length and
$h$ is the lattice step) with highest $\Delta S_i > \delta$.
Numerical experiments show that even small shares of such steps
drastically improve stability. More tests are presented
in~\cite{brownlee06}.

\begin{acknowledgments}
    The first author is grateful for a series of invaluable and encouraging discussions
    with Ilya Karlin and Shyam Chikatamarla.

    This work is supported by Engineering and Physical Sciences
    Research Council (EPSRC) grant number GR/S95572/01.
\end{acknowledgments}

\begin{figure*}
    \includegraphics[width=15.5cm]{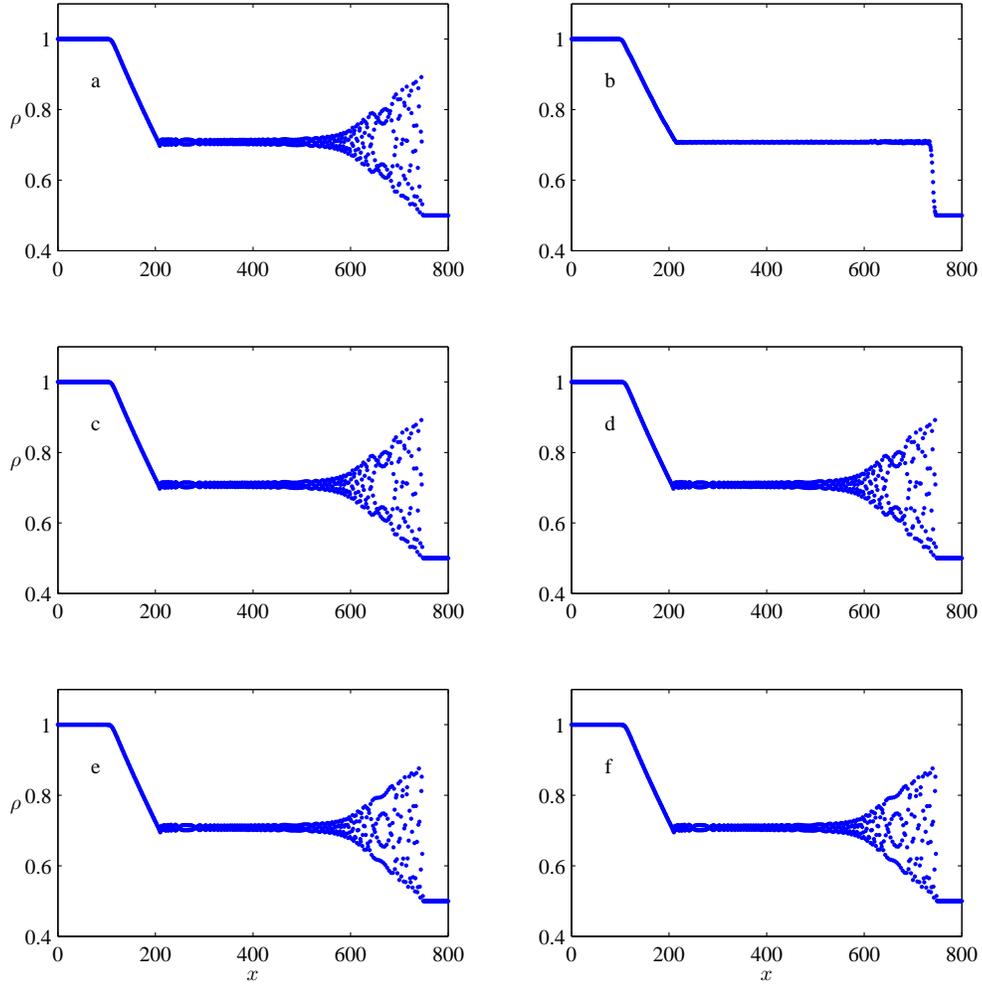}
    \caption{Density profile of the isothermal $1:2$ shock tube
    simulation after $500$ time steps using
    a) LBGK;
    b) LBGK-ES with $\delta=10^{-5}$;
    c) LBGK with regularisation provided by selecting the positivity bound $\alpha=\alpha_+$ if a population is
    predicted to become negative;
    d) LBGK with regularisation provided by selecting $\alpha=1$ if a population is predicted to become negative;
    e) ELBM with $\alpha=\alpha_+$ if no nontrivial root
    of~\eqref{entropy_estimate} exists;
    f) ELBM with $\alpha=1$ if no nontrivial root of~\eqref{entropy_estimate} exists.
    In this example, LBGK does not produce a negative population so the
    aforementioned regularisation procedures are redundant. Similarly,
    for ELBM in this example, the entropy estimate equation always has a
    nontrivial root.}
\end{figure*}

\begin{figure*}
    \includegraphics[width=15.5cm]{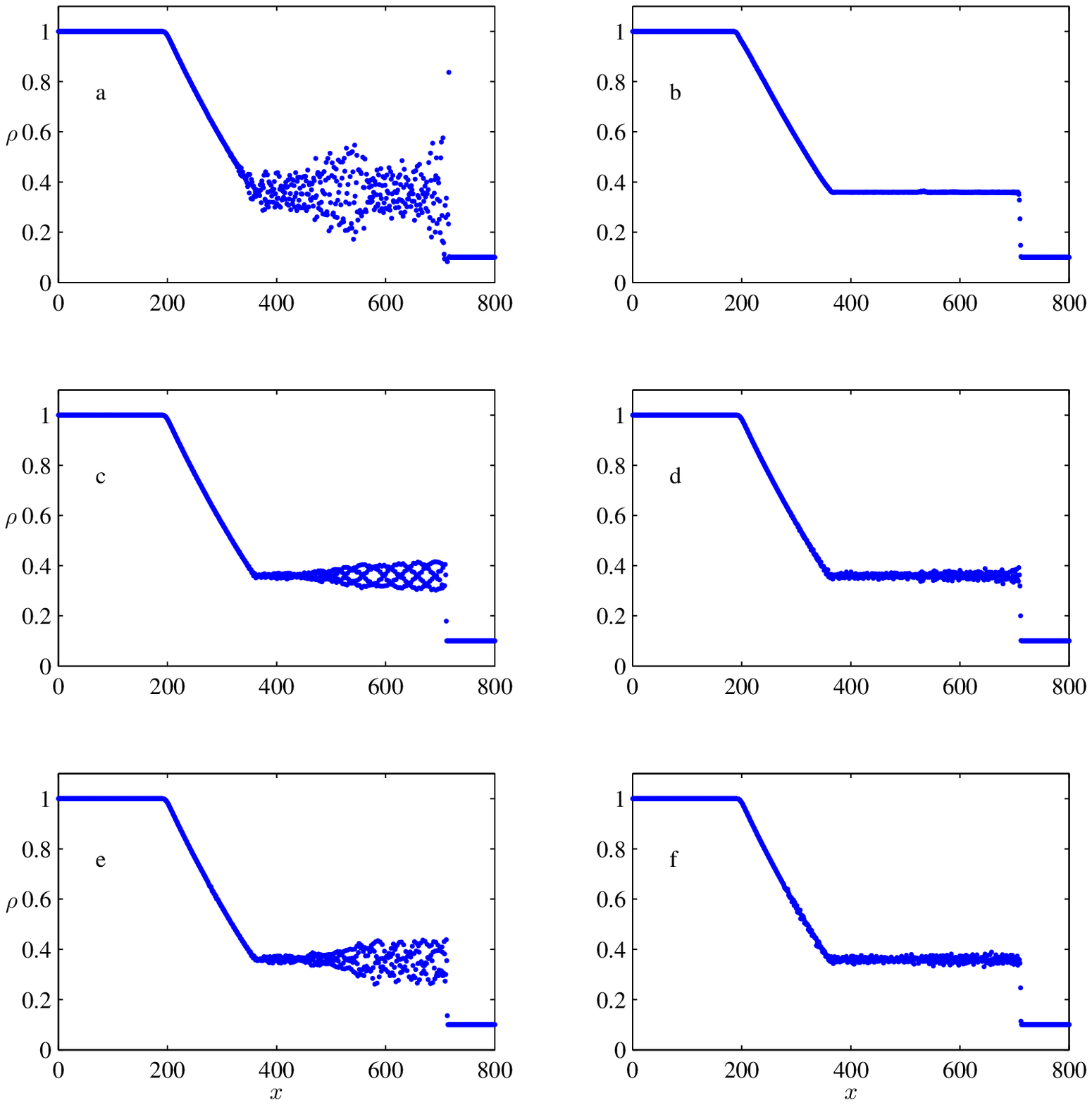}
    \caption{Density profile of the isothermal $1:10$ shock tube
    simulation after $350$ time steps using
    a) LBGK;
    b) LBGK-ES with $\delta=10^{-5}$;
    c) LBGK with regularisation provided by selecting the positivity bound $\alpha=\alpha_+$ if a population is
    predicted to become negative;
    d) LBGK with regularisation provided by selecting $\alpha=1$ if a population is predicted to become negative;
    e) ELBM with $\alpha=\alpha_+$ if no nontrivial root of~\eqref{entropy_estimate} exists;
    f) ELBM with $\alpha=1$ if no nontrivial root of~\eqref{entropy_estimate} exists.
    In this example, we acknowledge that the initial $1:10$ density discontinuity produces a shock which is outside of the hydrodynamic regime
    of the three-velocity model that we are using. However, this experiment is useful for testing the computational stability of the various
    LBMs because populations are capable of becoming negative.}\end{figure*}

\begin{figure*}
    \includegraphics[width=15.5cm]{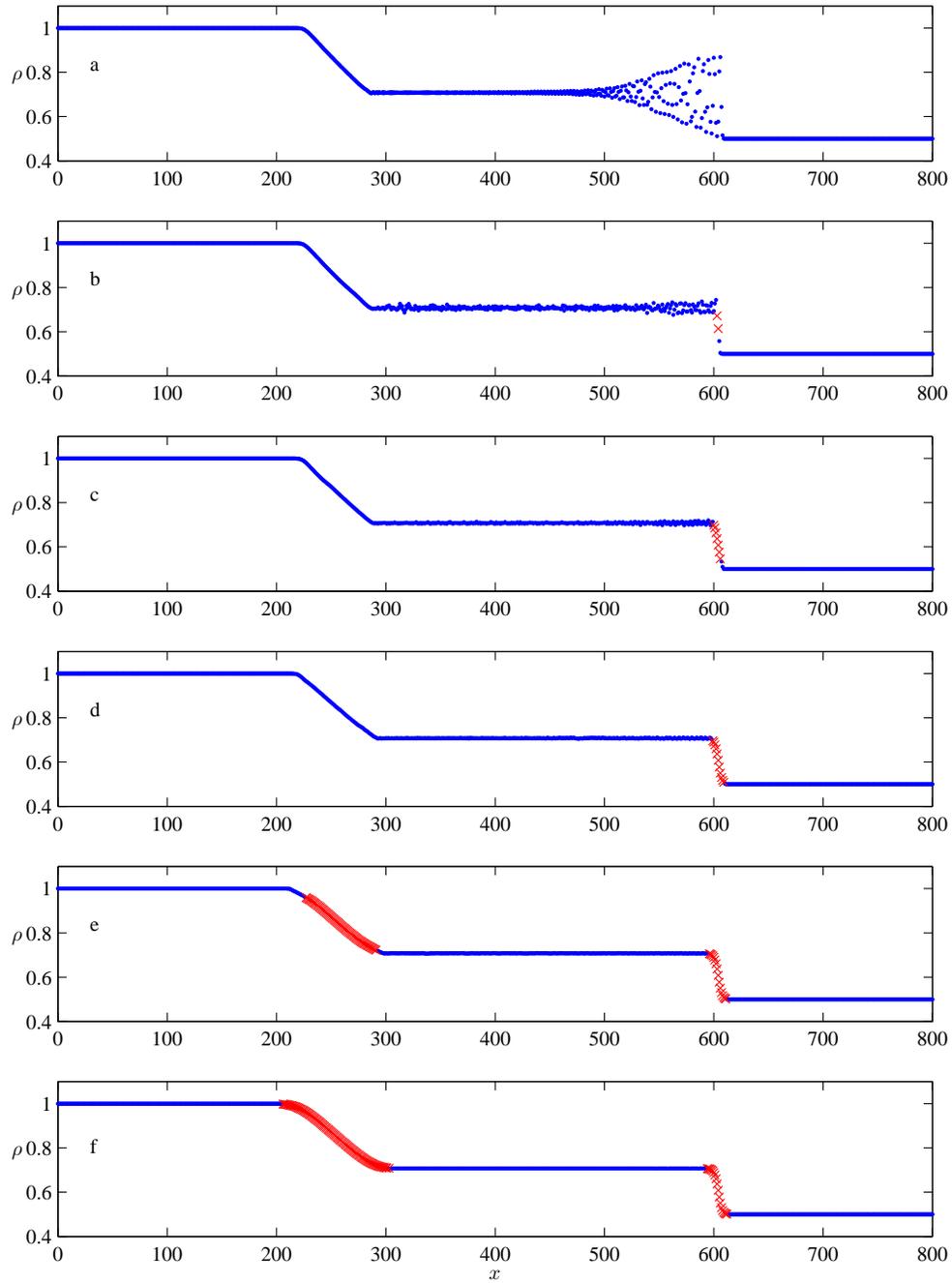}
    \caption{Density profile of the isothermal $1:2$ shock tube simulation
    after $300$ time steps using LBGK-ES with
    a) $\delta =10^{-2}$;
    b) $\delta =10^{-3}$;
    c) $\delta =10^{-4}$;
    d) $\delta =10^{-5}$;
    e) $\delta =10^{-6}$;
    f) $\delta =10^{-7}$.
    We are using crosses to indicate where an Ehrenfests' step is
    being used in the simulation. As we expect, the profile becomes progressively
    more smoothed as $\delta$ decreases.}
\end{figure*}

\begin{figure*}
    \includegraphics[width=15.5cm]{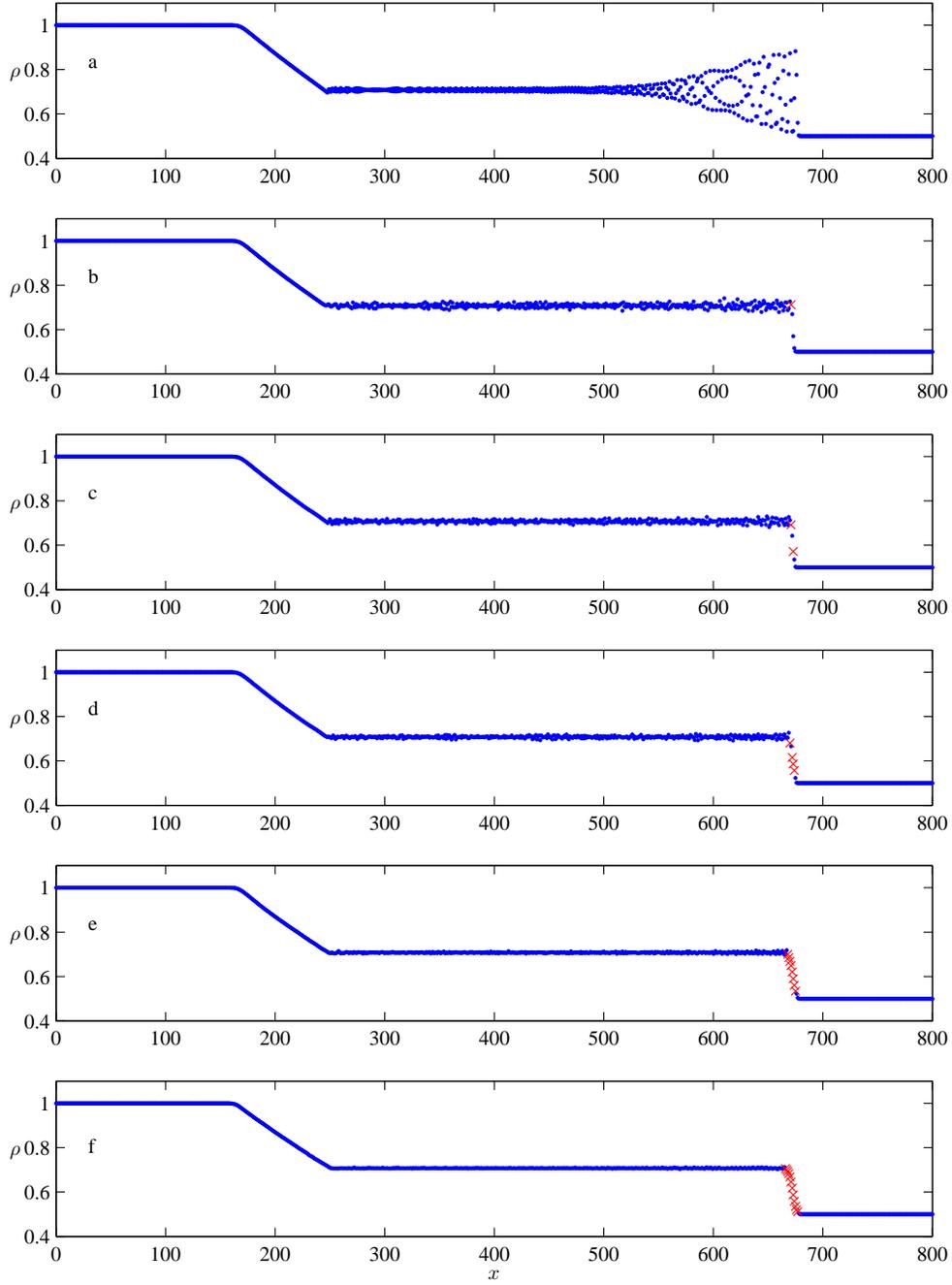}
    \caption{Density profile of the isothermal $1:2$ shock tube simulation
    after $400$ time steps using LBGK-ES with $\delta=10^{-5}$. Here, we
    are using the $(k,\delta)$-rule. This rule specifies that only
    the $k$ sites with the highest values of $\Delta S_i>\delta$ are
    accepted. The simulation shows the result for
    a) k=0;
    b) k=1;
    c) k=2;
    d) k=4;
    e) k=8;
    f) k=16.
    We are using crosses to indicate where an Ehrenfests' step is
    being used in the simulation. We observe that just a few Ehrenfests' steps greatly reduce spurious oscillations.}
\end{figure*}

\begin{figure*}
    \includegraphics[width=15.5cm]{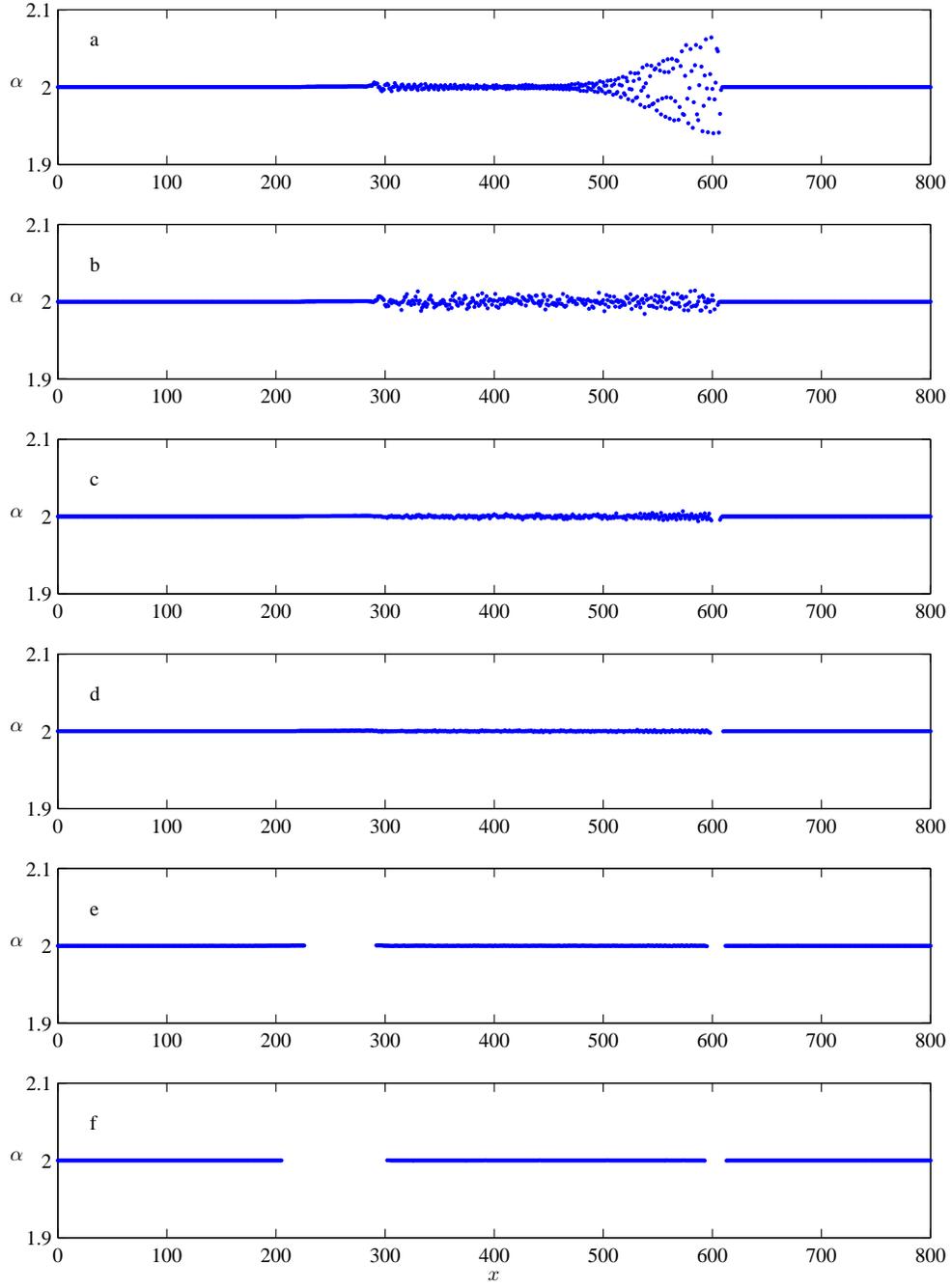}
    \caption{Profile of $\alpha\neq 1$ in the isothermal $1:2$ shock tube simulation
    after $300$ time steps using ELBM-ES with
    a) $\delta =10^{-2}$;
    b) $\delta =10^{-3}$;
    c) $\delta =10^{-4}$;
    d) $\delta =10^{-5}$;
    e) $\delta =10^{-6}$;
    f) $\delta =10^{-7}$.
    The gaps in the profiles occur precisely where $\alpha=1$ and
    have been removed so that the diminishing variation in
    $\alpha$ as $\delta$ decreases can be more readily observed.}
\end{figure*}

\begin{figure*}
    \includegraphics[width=18cm]{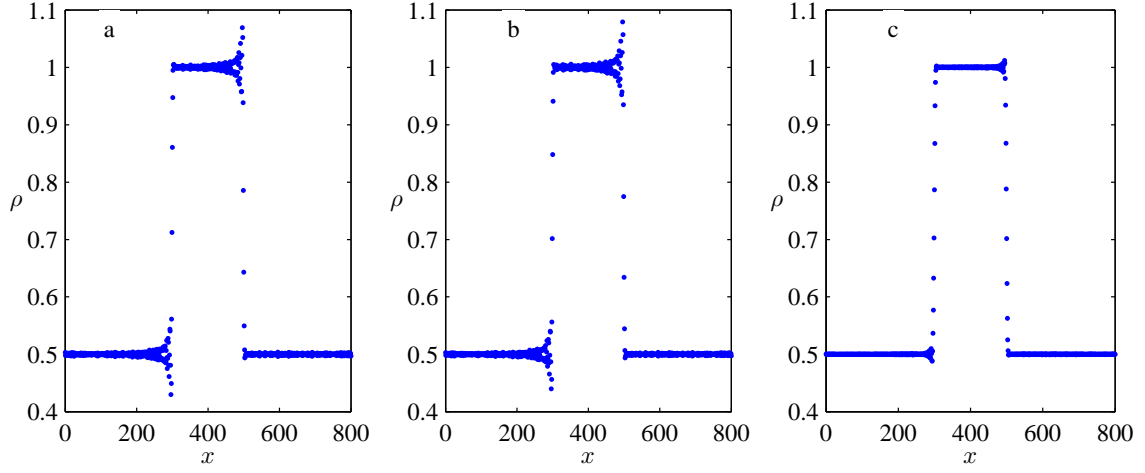}
 \caption{Advection of a square profile after $1000$ time
    steps using a) LBGK; b) ELBM and c) LBGK-ES with $\delta=10^{-3}$. All with
    $\beta=0.999$.}
\end{figure*}

\begin{figure*}
    \includegraphics[width=18cm]{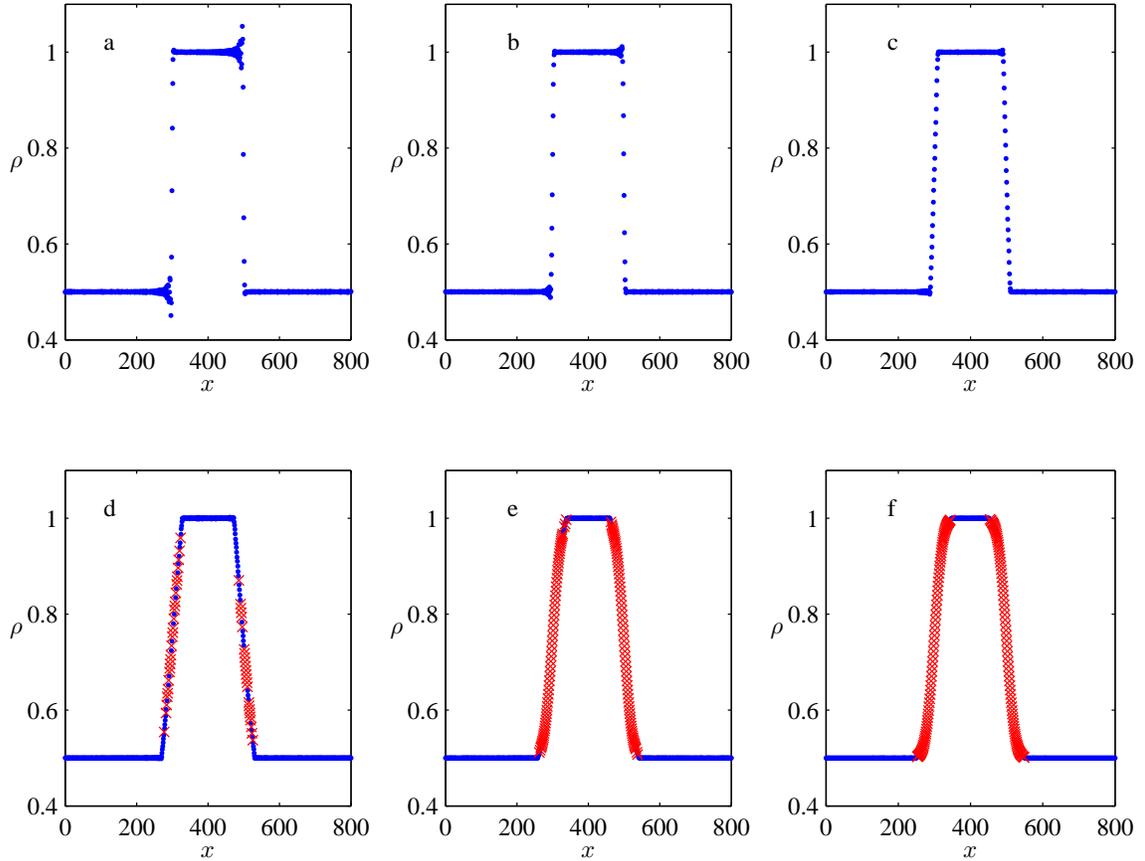}
 \caption{Advection of a square profile after $1000$ time
    steps using LBGK-ES ($\beta=0.999$) with
    a) $\delta =10^{-2}$;
    b) $\delta =10^{-3}$;
    c) $\delta =10^{-4}$;
    d) $\delta =10^{-5}$;
    e) $\delta =10^{-6}$;
    f) $\delta =10^{-7}$.
    We are using crosses to indicate where an Ehrenfests' step is
    being used in the simulation.}
\end{figure*}

\begin{figure*}
    \includegraphics[width=18cm]{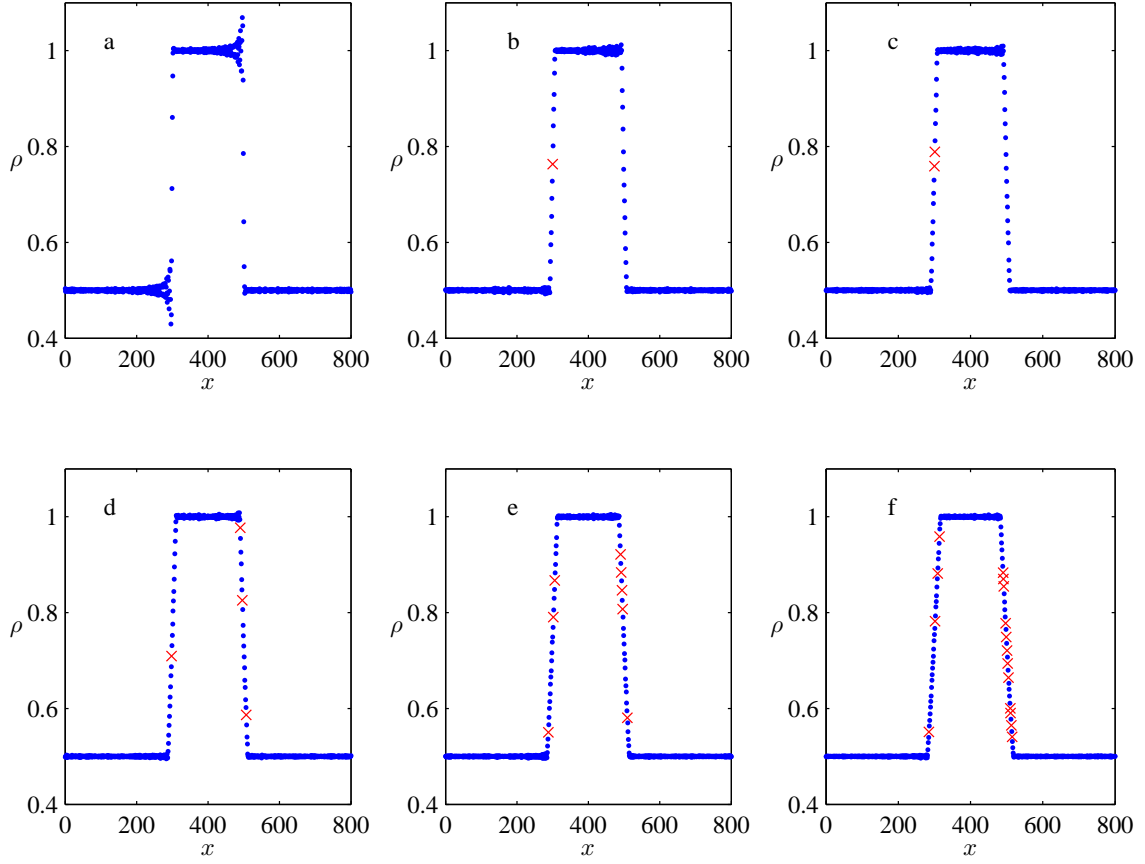}
 \caption{Advection of a square profile after $1000$ time
    steps using LBGK-ES ($\beta=0.999$) with $\delta=10^{-5}$.
    Here, we
    are using the $(k,\delta)$-rule. This rule specifies that only
    the $k$ sites with the highest values of $\Delta S_i>\delta$ are
    accepted. The simulation shows the result for
    a) k=0;
    b) k=1;
    c) k=2;
    d) k=4;
    e) k=8;
    f) k=16.
    We are using crosses to indicate where an Ehrenfests' step is
    being used in the simulation.}
\end{figure*}

\bibliographystyle{apsrev}

\end{document}